\documentclass[sigconf,nonacm]{acmart}
\usepackage{amsmath,mathtools}
\usepackage{enumitem}
\usepackage{xspace}
\usepackage{microtype}
\usepackage{booktabs}
\usepackage{tikz}
\usetikzlibrary{arrows.meta,positioning,calc,fit,backgrounds}
\tikzset{
  paper box/.style={draw,fill=white,rounded corners=2pt,align=center,line width=0.45pt},
  paper key/.style={paper box,fill=gray!12},
  paper frame/.style={draw=gray!70,rounded corners=3pt,line width=0.45pt},
  paper arrow/.style={->,>=Latex,line width=0.5pt},
  paper transform/.style={->,>=Latex,line width=0.75pt},
  paper secondary/.style={->,>=Latex,draw=gray!75,line width=0.45pt},
  paper guide/.style={dashed,draw=gray!60,line width=0.4pt}
}

\newcommand{\Log}{\mathsf{Log}}
\newcommand{\FAC}{\mathsf{FAC}}
\newcommand{\FAA}{\mathsf{FAA}}
\newcommand{\FAI}{\mathsf{FAI}}

\newcommand{\elog}{\mathsf{elog}}
\newcommand{\lelog}{\mathsf{lelog}}

\newcommand{\bits}{\mathsf{bits}}
\newcommand{\StableOrder}{\mathsf{StableOrder}}

\newcommand{\Tag}{\mathit{Tag}}
\newcommand{\Ticket}{\mathit{Ticket}}
\newcommand{\Head}{\mathit{Head}}
\newcommand{\Cert}{\mathit{Cert}}
\newcommand{\Pred}{\operatorname{Pred}}
\theoremstyle{remark}
\newtheorem{remark}{Remark}

\begin{document}

\title{Stabilizing Logs for Eventually Linearizable Shared Objects}

\author{Tong Che}
\affiliation{%
  \institution{NVIDIA Research}
  \country{USA}
}
\email{tongc@nvidia.com}

\begin{abstract}
Eventual linearizability permits a finite bad prefix but requires linearizable behavior thereafter.  We identify an eventually linearizable append-only log as the universal primitive for computable deterministic shared objects.  Conversely, every eventual log built from linearizable bases has a reachable configuration from which removing one finite prefix from every response yields a linearizable log.  This gives an exact target-side hierarchy equal to Herlihy's consensus hierarchy and a general upper bound for fixed finite pools of black-box eventual bases.  For matching lower bounds we introduce a compiler from stabilizing unordered predecessor certificates to an eventual log; its strongly-connected-component rule confines all inconsistent certificates to one finite prefix.  Self-describing fetch-and-add updates generate such certificates directly.  More surprisingly, an eventual fetch-and-increment counter also suffices for two processes: pre-announcements expose at most one unpublished ticket, and the returned rank determines whether that operation is a predecessor.  Thus the two standard long-lived counters have exact black-box values $\elog(\FAI)=\elog(\FAA)=2$, while solo-explainable one-shot types remain at level $1$.
\end{abstract}


\maketitle

\section{Introduction}

Linearizability gives each concurrent object operation the illusion of taking effect at a single instant between its invocation and response \cite{HerlihyWing1990}.  Wait-freedom requires every operation to finish regardless of the speeds of other processes \cite{Herlihy1991}.  The interaction between these two requirements is measured by Herlihy's consensus hierarchy: a type $T$ has consensus number $c(T)$ if $T$ and registers can implement consensus for $c(T)$ processes, but not for $c(T)+1$ processes \cite{Herlihy1991}.

Eventual linearizability weakens the safety side.  Histories must be weakly consistent, and there must exist a finite point after which the history admits a linearization that respects real-time order and return values in the suffix \cite{Serafini2010,GuerraouiRuppert2014}.  Guerraoui and Ruppert showed that this weakening creates a paradox: consensus and test-and-set become easy to implement eventually, while fetch-and-increment remains as hard as its linearizable counterpart when implemented from linearizable base objects \cite{GuerraouiRuppert2014}.

We ask whether eventual linearizability admits a Herlihy-style hierarchy, and we identify the right universal primitive for it.  Consensus is not that primitive, since it collapses under eventual linearizability.  Instead we use a long-lived operation log.  Fetch-and-cons is one presentation of such a log: $\FAC(v)$ returns the current list and then adds $v$.  An eventually linearizable log is universal for eventual objects because it supplies the order in which operations should eventually be replayed.  Basing a hierarchy on this primitive requires care at two points, which we treat explicitly.

First, the stabilization cut must remove the complete log state at the cut.  If an operation $o_0=\FAC(x)$ returns $L_0$, then after $o_0$ the abstract log state is $L_0$ followed by $x$, so removing only $L_0$ would leave $x$ visible to the first operation of the supposedly fresh object.  Second, the stable-configuration argument initializes the derived implementation at a reachable configuration of the original one; this is sound for lower-bound arguments only when the relevant base-object types are robust to reachable initial states, or when the theorem is stated explicitly as a shifted-initial-state theorem.  We make both points precise and discharge them.

\paragraph{Contributions.}
\begin{enumerate}[leftmargin=2em]
    \item \emph{Stabilizing logs.}  An eventual log is universal, while an eventual log built from linearizable bases can be shifted and prefix-quotiented into a linearizable log.  Consequently $\lelog(T)=c(T)$ for state-robust $T$, and finite-pool black-box bases satisfy $\elog(T)\le c(T)$.
    \item \emph{A predecessor-certificate compiler.}  Safe unordered certificates that stabilize to one common finite-past order are compiled into ordered log prefixes; delayed publication and every inconsistent certificate are absorbed into one finite SCC prefix.
    \item \emph{Exact eventual counters.}  Self-describing FAA updates instantiate the compiler directly, while a new rank-lifting construction reconstructs predecessor identities from an eventual FAI response and one possible ticket hole.  Hence $\elog(\FAI)=\elog(\FAA)=2$.
\end{enumerate}

The message is twofold.  Target-side eventuality costs nothing under the standard class-level convention.  Base-side eventuality needs finite-contamination arguments, but both identity-bearing updates and apparently weaker ranks can generate a stable log.

\section{Related work}

\paragraph{Consistency conditions.}
Linearizability is the standard correctness condition for concurrent objects \cite{HerlihyWing1990}; it strengthens sequential consistency \cite{Lamport1979} by additionally requiring real-time order to be respected.  The idea of representing an object by an agreed sequence of operations originates in the replicated state-machine approach \cite{Lamport1978,Schneider1990,Lamport1998}, which our log-based universal construction mirrors in shared memory.  Standard background on these models can be found in \cite{AttiyaWelch2004}.

\paragraph{Wait-free hierarchy.}
Wait-freedom and the consensus hierarchy were introduced by Herlihy \cite{Herlihy1991}, building on the asynchronous consensus impossibility of Fischer, Lynch, and Paterson \cite{FLP1985} and of Loui and Abu-Amara \cite{LouiAbuAmara1987}.  Herlihy's universal construction shows that any object with sufficient consensus power is universal, and Plotkin's sticky bits give a further universal primitive \cite{Plotkin1989}.  Our universality theorem is the eventual-linearizability analogue: a single eventually linearizable log replaces the tower of consensus instances.

\paragraph{Limits of the consensus number.}
The consensus number classifies universal power, not every pairwise reduction.  Jayanti studied robustness under combinations of types \cite{Jayanti1993,Jayanti1997}, and later separations showed that equal-numbered deterministic types can have different power \cite{AfekEllenGafni2016,Daian2018,Zhu2025}.  Our hierarchy likewise addresses universal implementability; the counter results require additional structure beyond the scalar upper bound.

\paragraph{Relaxed and eventual consistency.}
Weakening consistency to gain availability or concurrency is widespread.  Eventual consistency underlies large-scale replicated stores \cite{Terry1995,Vogels2009}, and conflict-free replicated data types make convergence a type-level guarantee \cite{Shapiro2011}.  In shared memory, quasi-linearizability \cite{Afek2010} and quantitative relaxation \cite{Henzinger2013} bound how far responses may deviate from a linearization.  Eventual linearizability differs from all of these: it keeps the object's sequential specification intact and bounds only the \emph{duration} of inconsistency, requiring exactly linearizable behavior after a finite prefix.

\paragraph{Eventual linearizability and stabilization.}
Eventual linearizability was formalized by Serafini et al.\ \cite{Serafini2010}, whose eventual-agreement layer also orders operations for generic object construction in a message-passing setting.  We study wait-free shared-memory reductions between object types.  Guerraoui and Ruppert \cite{GuerraouiRuppert2014} proved both the easiness of eventual consensus and test-and-set and the hardness of eventual fetch-and-increment from linearizable bases.  Our restoration theorem generalizes their stable-configuration argument from counters to logs.  Our certificate compiler is related to self-stabilization \cite{Dijkstra1974,Dolev2000}: it tolerates arbitrary but finite early certificate anomalies without knowing when they end.

\section{Model and definitions}

We use the standard asynchronous shared-memory model with a fixed set $P=\{p_1,\ldots,p_n\}$ of deterministic processes.  A history is a sequence of invocation and response events.  It is well formed if every process history alternates invocations and matching responses, possibly ending with one pending invocation.  An operation is complete if its response appears.

An object type $T$ is a deterministic labelled transition system with states $S$, initial states $S_0$, invocations $I$, responses $V$, and transition function $\delta:S\times I\to V\times S$.  A sequential history is legal for $T$ if it follows $\delta$ from some state in $S_0$.  A type is \emph{computable} when states, invocations, and responses have effective encodings, it has a distinguished effectively representable initial state $s_0\in S_0$, and $\delta$ is total computable.  A concurrent history is linearizable if it can be completed and reordered into a legal sequential history preserving per-process order and real-time precedence between non-overlapping operations \cite{HerlihyWing1990}.  All base objects used in reductions provide wait-free operations; eventual linearizability is their safety condition, not a progress condition.

Unless a cardinality restriction is stated explicitly, we use the standard class-level resource convention of the consensus hierarchy: an implementation may draw from an unbounded, statically named supply of base objects and registers.  Every finite execution accesses only finitely many of them.  Processes are Turing-computable; finite integers, finite sets, and register addresses have unbounded representations, and local work is charged in the usual step model.  This is the convention used by Herlihy's consensus number and universal construction.

\begin{definition}[Weak consistency]
A well-formed history $H$ of a single object is weakly consistent if, for every complete operation $op$ by process $p$, there exists a legal sequential history $S_{op}$ such that: (i) every operation in $S_{op}$ was invoked in $H$ before $op$ responded; (ii) $S_{op}$ contains all complete operations by $p$ that precede $op$ in $H$; and (iii) the last operation of $S_{op}$ is $op$ with the same response as in $H$.
\end{definition}

\begin{definition}[$t$-linearizability and eventual linearizability]
Let $H^{\ge t}$ be the suffix of $H$ obtained by deleting the first $t$ events.  A legal sequential history $S$ is a $t$-linearization of $H$ if: (i) every operation in $S$ is invoked in $H$; (ii) every complete operation in $H$ appears complete in $S$; (iii) if $op_1$ responds before $op_2$ is invoked, both events are in $H^{\ge t}$, and $op_2$ appears in $S$, then $op_1$ precedes $op_2$ in $S$; and (iv) every operation whose response occurs in $H^{\ge t}$ has in $S$ the same response as in $H$.  A history is eventually linearizable if it is weakly consistent and $t$-linearizable for some finite $t$.
\end{definition}

Different histories may have different stabilization points $t$.  The definition is a history property, not an online condition; an implementation is eventually linearizable when every history it produces has the property.

\subsection{Logs and fetch-and-cons}

We use an append-only log because it removes notational ambiguity.  A log state is a finite sequence $L$ of unique tags.  The operation $\Log.append(x)$ returns the current sequence $L$ and changes the state to $L\cdot x$.  Fetch-and-cons is equivalent up to reversing the sequence; all algorithms and lower bounds translate immediately.  In an implementation, each high-level operation uses a tag
\[
    x=(p_i,k,op,args),
\]
where $k$ is a local sequence number.  Tags are unique even if two operations have the same payload.

If $A$ and $B$ are sequences and $A$ is a prefix of $B$, write $B\ominus A$ for the unique suffix $C$ such that $B=A\cdot C$.  Prefix quotients are the formal version of ``subtracting the finite bad prefix.''  This definition is intentionally sequence-based rather than set-based: duplicates and order both matter for logs.

When we write $\Log_n$, we mean the canonical $n$-process log, with operations indexed by the fixed process set $P=\{p_1,\ldots,p_n\}$.  Consensus numbers for such process-indexed objects are understood with this arity convention.  Thus a linearizable $\Log_n$ solves $n$-process consensus by deciding the value in the first log position, and conversely $n$-consensus implements $\Log_n$ by Herlihy's universal construction; hence $c(\Log_n)=n$.

\begin{definition}[Eventual log numbers]
Let $c(T)$ be Herlihy's consensus number.  Define $\lelog(T)$ to be the largest $n$ (or $\infty$) such that wait-free linearizable objects of type $T$, under the standard class-level supply convention, together with linearizable registers, can wait-free implement an eventually linearizable $n$-process log.  Define $\elog(T)$ to be the largest $n$ (or $\infty$) such that a fixed finite collection of wait-free eventually linearizable objects of type $T$, together with linearizable registers, can wait-free implement an eventually linearizable $n$-process log, where the implementation must be correct for every history of the base objects satisfying their eventual-linearizability specification.
\end{definition}

The asymmetry is deliberate.  The first number follows the same class-level convention as $c(T)$ and isolates the effect of weakening the target object.  For black-box eventual bases, independent per-object guarantees provide no common stabilization point for an unbounded supply: an implementation could keep accessing fresh objects whose finite bad prefixes occur later and later.  The second number therefore fixes a finite pool.  Thus $\lelog$ and $\elog$ answer distinct resource questions rather than differing only in a consistency bit.  For upper bounds, linearizable histories of that finite pool are still a subset of its eventually linearizable histories.

\begin{definition}[State-robust type]
A type $T$ is state-robust for $n$ processes if allowing any finite subset of a standard supply of $T$ objects to start in fixed states reachable by finite legal sequential executions does not increase its $n$-process consensus power.  Equivalently, if such a supply---with only finitely many copies shifted to reachable states---and registers implements $n$-consensus, then a standard supply of $T$ objects and registers also implements $n$-consensus.
\end{definition}

We call $T$ \emph{state-robust} without qualification when it is state-robust for every $n$; the finitely many shifted states are fixed known constants.  Registers, unbounded $\FAI$ and $\FAA$ counters, and fetch-and-cons logs are state-robust: a fixed register value, counter offset, or log prefix can be absorbed by a fixed translation.  We use state robustness only for these named types.

\section{An eventual log is universal}

The classical universal construction orders operations by a sequence of consensus instances.  For eventual linearizability, a single eventual log suffices.  Each process announces its operation by appending a unique tag to the log, replays the returned prefix on a local copy, and then applies its own operation.

\begin{quote}
\textbf{Universal construction from a log.}
Process $p_i$ maintains a local sequence number $k_i$.
To perform operation $op(args)$ on a deterministic type $T$:
\begin{enumerate}[leftmargin=2em]
    \item Let $x=(p_i,k_i,op,args)$ and increment $k_i$.
    \item Invoke $L_x:=\Log.append(x)$.
    \item Initialize a local copy $q$ of $T$ to its initial state.
    \item Replay, in order, the operation encoded by every tag in $L_x$ on $q$.
    \item Apply $op(args)$ to $q$ and return the response.
\end{enumerate}
\end{quote}

The construction is wait-free if the log is wait-free and $T$ has computable transitions.  The replay cost grows with the log prefix; this is the standard unbounded-space universal-construction cost and is irrelevant for computability.

\begin{theorem}[Eventual log universality]
For every computable deterministic type $T$, a wait-free eventually linearizable $n$-process log and registers wait-free implement an eventually linearizable $n$-process object of type $T$.
\end{theorem}

\begin{proof}
Let $H$ be a history produced by the construction, and let $G$ be the corresponding history of operations on the log.  Since the log is eventually linearizable, $G$ is weakly consistent and has a $t_L$-linearization $S_L$ for some $t_L$.

First we prove weak consistency of $H$.  Consider a complete high-level operation $op_x$ with tag $x$ and log response $L_x$.  By weak consistency of the log, there is a legal sequential log history $E_x$ explaining the response $L_x$ of $\Log.append(x)$.  In any legal sequential log history, the response to $\Log.append(x)$ is exactly the sequence of tags that precede $x$.  Therefore, replaying the operations encoded by $L_x$ and then applying the operation encoded by $x$ gives precisely the response returned by the construction.  The mapped sequential history contains only operations whose tags were appended before $op_x$ returned, and it contains every earlier operation by the same process because a process appends its tags in local order and the explanation for $x$ must include that process's previous appends.  Thus it is an explanation history for $op_x$.

Now we prove eventual suffix linearizability.  Let $B$ be the finite set of high-level operations whose log response occurs among the first $t_L$ events of $G$.  Choose $t$ in $H$ after every operation in $B$ that completes in $H$ has responded.  We construct a $t$-linearization $S$ of $H$ by taking the $t_L$-linearization $S_L$ of the log and replacing each appended tag by the corresponding high-level operation.  If a high-level operation has appended its tag but is still pending in $H$, it may be completed in $S$ with the response dictated by the replay, as allowed in the completion used for linearizability.

Every complete operation in $H$ completed its log append, hence appears in $S_L$ and therefore appears in $S$.  If $op_x$ responds before $op_y$ is invoked in $H$ and both events are after $t$, then the response of $\Log.append(x)$ precedes the invocation of $\Log.append(y)$ after $t_L$ in $G$.  Since $S_L$ respects real-time order after $t_L$, $x$ precedes $y$ in $S_L$, so $op_x$ precedes $op_y$ in $S$.

It remains to check responses after $t$.  Let $op_x$ respond after $t$.  Its log response is after $t_L$, so $S_L$ gives the same log response $L_x$ as the concurrent execution.  In a legal sequential log history, $L_x$ is exactly the sequence of tags preceding $x$.  The high-level operation $op_x$ replays exactly those tags and then applies its own operation; therefore its response in $S$ is the response returned in $H$.  Thus $S$ is a $t$-linearization.  Together with weak consistency, $H$ is eventually linearizable.
\end{proof}

\section{Stable configurations and cutting a log}\label{sec:stable}

This section proves that target-side eventuality does not make logs easier to implement from linearizable base objects.  The proof follows the stable-node strategy of Guerraoui and Ruppert for fetch-and-increment, but the cut is a sequence prefix rather than an integer offset.

Consider an implementation $A$ of an $n$-process log from linearizable base objects.  Its execution tree ranges over every schedule and every sequence of fresh tagged append inputs; each edge is one implementation step.  A node $C$ corresponds to a finite execution prefix $\alpha_C$, and $d(C)$ is its tree depth.  We call $C$ \emph{stable} if every execution extending $\alpha_C$ is $d(C)$-linearizable as a log history.  The integer $d(C)$ is used as an event-index cutoff; it need not equal the number of high-level events already produced.

We use two simple facts.  First, $t$-linearizability is monotone: if a history is $t$-linearizable, it is also $t'$-linearizable for every $t'\ge t$.  Second, for log histories produced by an eventually linearizable implementation, $t$-linearizability is closed under limits: if every finite prefix of an infinite history is $t$-linearizable, then the infinite history is $t$-linearizable.  The second fact is the log analogue of the fetch-and-increment lemma of Guerraoui and Ruppert.  The limit- and prefix-closure lemmas below are the log analogues of the compactness arguments of \cite{GuerraouiRuppert2014}; the stable-node and good-cut lemmas that follow adapt their stable-configuration method from an integer offset to a sequence prefix.

\begin{lemma}[Limit closure for logs]\label{lem:limit}
Let $\alpha$ be an infinite history of an eventually linearizable log implementation.  If every finite prefix of $\alpha$ is $t$-linearizable, then $\alpha$ is $t$-linearizable.
\end{lemma}

\begin{figure*}[t]
\centering
\begin{tikzpicture}[font=\footnotesize,>=Latex]
  \node[font=\small] at (4.15,1.65) {\textbf{Compatible finite orders}};
  \node[font=\small] at (11.55,1.65) {\textbf{Coherent limit}};
  \node[anchor=east] at (0.65,1.0) {level $M_1$};
  \node[anchor=east] at (0.65,0.0) {level $M_2$};
  \node[anchor=east] at (0.65,-1.0) {level $M_3$};

  \node[paper key,minimum width=1.45cm,minimum height=0.58cm] (a1) at (2.05,1.0) {$\sigma_1$};
  \node[paper box,minimum width=1.45cm,minimum height=0.58cm] (a2) at (4.25,1.0) {$\sigma'_1$};
  \node[paper key,minimum width=1.45cm,minimum height=0.58cm] (b1) at (2.05,0.0) {$\sigma_2$};
  \node[paper box,minimum width=1.45cm,minimum height=0.58cm] (b2) at (4.25,0.0) {$\sigma'_2$};
  \node[paper box,minimum width=1.45cm,minimum height=0.58cm] (b3) at (6.45,0.0) {$\sigma''_2$};
  \node[paper key,minimum width=1.45cm,minimum height=0.58cm] (c1) at (2.05,-1.0) {$\sigma_3$};
  \node[paper box,minimum width=1.45cm,minimum height=0.58cm] (c2) at (4.25,-1.0) {$\sigma'_3$};
  \node[paper box,minimum width=1.45cm,minimum height=0.58cm] (c3) at (6.45,-1.0) {$\sigma''_3$};

  \draw[paper transform] (c1) -- (b1);
  \draw[paper transform] (b1) -- node[right]{restrict} (a1);
  \draw[paper secondary] (c2) -- (b2);
  \draw[paper secondary] (c3) -- (b2);
  \draw[paper secondary] (b2) -- (a2);
  \draw[paper secondary] (b3) -- (a2);
  \node at (2.05,-1.65) {$\vdots$};
  \node[align=center] at (4.15,-1.65) {highlighted edges form one coherent restriction path};

  \draw[paper transform] (7.25,0) -- node[above]{K\"onig's lemma} (9.15,0);
  \node[paper key,minimum width=4.4cm,minimum height=1.12cm] (limit) at (11.55,0)
    {coherent family $(\sigma_k)_{k\ge1}$\\$\Longrightarrow$ total order on $\bigcup_k M_k$};
  \node[align=center] at (11.55,-1.65) {returned prefixes bound every operation's predecessor set};
\end{tikzpicture}
\caption{Compactness in Lemma~\ref{lem:limit}.  Each level contains the finite orders compatible with the first $k$ fixed responses; restriction edges form a finitely branching tree.  K\"onig's lemma selects a coherent path, whose union is the required order.}
\Description{A finitely branching tree of compatible finite orders has a highlighted infinite path leading to one coherent order on their union.}
\label{fig:limit-tree}
\end{figure*}

\begin{proof}
Let $R=\{r_1,r_2,\ldots\}$ be the complete operations whose responses occur after $t$, listed in response order, and let $E$ be the finite set of complete operations whose responses occur before $t$.  For each $k$, let $M_k$ contain $E$, the operations $r_1,\ldots,r_k$, and every operation whose tag occurs in one of the returned prefixes of $r_1,\ldots,r_k$.  The set $M_k$ is finite.  A node at level $k$ is a total order of $M_k$ such that, for every $i\le k$, the operations preceding $r_i$ are exactly the operations named in the prefix returned by $r_i$, in the returned order, and the order respects real-time precedence after $t$ among operations in $M_k$.  Edges are restrictions from level $k+1$ to level $k$.

Every level is nonempty.  Indeed, take a finite prefix of $\alpha$ containing the responses of $r_1,\ldots,r_k$ and the invocations of all operations whose tags occur in their returned prefixes.  By assumption this finite prefix has a $t$-linearization.  Restricting that linearization to $M_k$ gives a level-$k$ node: no operation outside the returned prefix of $r_i$ can precede $r_i$, since a legal log would then include its tag in $r_i$'s response.  The tree is finitely branching, so Koenig's lemma gives a coherent family of finite orders and hence a total order $<$ on the union $M=\bigcup_k M_k$ satisfying all finite constraints.

It remains only to check that this total order is a sequential history, i.e. that no complete operation is pushed behind infinitely many later operations.  The eventual-linearizability of the implementation supplies a $t'$-linearization $S'$ of $\alpha$.  If $R$ is infinite, every operation in $E$ appears at a finite position in $S'$, and therefore appears in the actual returned prefix of every sufficiently late operation of $R$ whose response is after $t'$.  Hence each operation of $E$ is named in some returned prefix used in the construction above.  If an operation is some $r_i$, its predecessors under $<$ are exactly the finite returned prefix of $r_i$; if it is named in the returned prefix of some $r_i$, then all of its predecessors are contained in that same finite prefix.  Thus every operation in $M$ has finitely many predecessors.  When $R$ is finite, the same conclusion is immediate because only finitely many complete operations must be placed.

Enumerating $M$ in any order extending $<$ therefore gives a legal sequential log history.  It contains every complete operation of $\alpha$: those before $t$ are in $E$, and those after $t$ are in $R$.  For each operation responding after $t$, the construction makes its predecessor sequence exactly its returned prefix, and the order respects real-time precedence after $t$.  Thus the resulting sequential history is a $t$-linearization of $\alpha$.
\end{proof}

\begin{lemma}[Prefix closure for logs]\label{lem:prefix-closure}
If a log history $H$ is $t$-linearizable, then every finite prefix $H'$ of $H$ is $t$-linearizable.
\end{lemma}

\begin{proof}
Let $S$ be a $t$-linearization of $H$.  Start with the subsequence of $S$ consisting of operations invoked in $H'$.  If an operation $op$ responds in $H'$ after $t$, then its response is fixed by the definition of $t$-linearizability.  In a legal log history this response is exactly the sequence of tags preceding $op$ in $S$.  Any such predecessor $q$ appears in $S$ and must have been invoked before $op$ responded: if it were invoked after that response, both events would lie after $t$, and clause~(iii), with $q\in S$, would force $op$ to precede $q$.  Hence all predecessors required by the fixed response of $op$ are invoked in $H'$.  Therefore removing operations not invoked in $H'$ does not remove any predecessor required by a post-$t$ response.  Operations whose responses occur before $t$, and pending operations, may be completed with the responses determined by their positions in the restricted sequence.  The resulting legal sequential log history contains all complete operations of $H'$, respects real-time order after $t$ because $S$ does, and gives every operation responding after $t$ the same response as in $H'$.  Thus it is a $t$-linearization of $H'$.
\end{proof}

\begin{lemma}[Existence of a stable node]\label{lem:stable-node}
Every wait-free eventually linearizable log implementation has a stable node in the execution tree in which all processes repeatedly append fresh tags.
\end{lemma}

\begin{proof}
Assume no stable node exists.  We build nonempty finite paths $p_0,p_1,p_2,\ldots$ such that, for $\ell_i=|p_0p_1\cdots p_i|$, the prefix through $p_i$ is not $\ell_{i-1}$-linearizable for every $i\ge1$.  Given the prefix through $p_{i-1}$, its endpoint is not stable, so some extension is not $\ell_{i-1}$-linearizable.  If it is finite, take it; if it is infinite, the contrapositive of Lemma~\ref{lem:limit} supplies a bad finite prefix.  Choose the additional nonempty path as $p_i$.  Since paths are nonempty step sequences, $\ell_i\to\infty$.

Let $\pi=p_0p_1p_2\cdots$ be the infinite execution.  Since the implementation is eventually linearizable, $\pi$ is $t$-linearizable for some finite $t$.  Choose $i$ with $\ell_{i-1}>t$.  By monotonicity, $\pi$ is $\ell_{i-1}$-linearizable, so by Lemma~\ref{lem:prefix-closure} every finite prefix of $\pi$ is $\ell_{i-1}$-linearizable, contradicting the construction of $p_i$.
\end{proof}

\begin{lemma}[Good cut]\label{lem:good-cut}
Let $C$ be a stable node and let $t=d(C)$.  There is an extension $\alpha_0$ from the root to a configuration $C_0$ and a completed operation $o_0$ in $\alpha_0$ such that: (i) $o_0$ responds after event $t$; (ii) every operation invoked before $C_0$ is complete; and (iii) in every $t$-linearization of every continuation of $\alpha_0$, all operations invoked before $C_0$ precede every operation invoked after $C_0$ that appears in the linearization.
\end{lemma}

\begin{proof}
From $C$, first let each process run solo until its current operation, if any, completes; wait-freedom reaches an idle configuration $I$.  Then let one process perform append operations alone forever, and fix a $t$-linearization $S$ of this infinite solo extension, which exists because $C$ is stable.  The operations invoked before $I$, together with the finitely many solo operations having an event among the first $t$ high-level events, form a finite set whose members all occur at finite positions in $S$.  The remaining solo tail is ordered in $S$ by suffix real time.  Choose $o_0=\Log.append(x_0)$ sufficiently far into this tail that every operation in the finite set and every earlier tail operation precedes $o_0$ in $S$.

Because $o_0$ responds after $t$, its response in $S$ is its actual response.  In a legal log history this response is exactly the sequence of tags preceding $o_0$, and by the choice of $o_0$ those are precisely the tags of the operations invoked before $o_0$.  Let $C_0$ be the configuration immediately after $o_0$ completes, and let $P_0$ be the returned prefix followed by $x_0$; then $P_0$ is exactly the abstract log state containing all operations invoked before $C_0$.

Now consider any continuation of $\alpha_0$, where $\alpha_0$ is the finite execution ending at $C_0$, and any $t$-linearization of that continuation.  Since $o_0$ responds after $t$, this linearization must give $o_0$ the same actual response, so the operations in the returned prefix again precede $o_0$.  If an operation invoked after $C_0$ appears in the linearization, then the response of $o_0$ precedes that invocation and both events are after $t$; clause~(iii) therefore places $o_0$ before it.  Hence every operation invoked before $C_0$ precedes every later operation that appears in the linearization.
\end{proof}

\begin{figure*}[t]
\centering
\begin{tikzpicture}[font=\footnotesize,>=Latex]
  \node[anchor=east,font=\small] at (-0.15,1.05) {\textbf{ordering}};
  \node[anchor=east,font=\small] at (-0.15,-0.55) {\textbf{execution}};

  \node[paper key,minimum width=3.1cm,minimum height=0.9cm] (P0) at (4.8,1.05)
    {$P_0=L_0\cdot x_0$\\all pre-cut operations};
  \node[paper box,minimum width=5.0cm,minimum height=0.9cm] (post) at (10.3,1.05)
    {operations after $C_0$\\included in the linearization};
  \draw[paper transform] (P0) -- node[above]{precedes} (post);

  \node[paper box,minimum width=1.4cm,minimum height=0.82cm] (C) at (0.85,-0.55) {stable $C$};
  \node[paper key,minimum width=3.0cm,minimum height=0.82cm] (alpha) at (3.35,-0.55)
    {finite $\alpha_0$\\complete pending work; solo appends};
  \node[paper box,minimum width=1.7cm,minimum height=0.82cm] (o0) at (6.0,-0.55)
    {$o_0$\\$\Log.append(x_0)$};
  \draw[paper secondary] (C) -- (alpha);
  \draw[paper secondary] (alpha) -- (o0);

  \draw[paper guide] (7.2,-1.25) -- (7.2,0.15);
  \node at (7.2,-1.42) {$C_0$};
  \draw[paper secondary] (o0) -- (7.2,-0.55);

  \node[paper box,minimum width=5.0cm,minimum height=0.82cm] (suffix) at (10.3,-0.55)
    {arbitrary continuation $\beta$\\fresh appends after the cut};
  \draw[paper secondary] (7.2,-0.55) -- (suffix.west);
  \draw[paper secondary] (suffix.east) -- node[right] {$\cdots$} (13.6,-0.55);
\end{tikzpicture}
\caption{The good-cut lemma.  The execution reaches $C_0$ after a sufficiently late solo append $o_0$; every post-cut operation included in a continuation's linearization follows the full state $P_0=L_0\cdot x_0$.}
\Description{A stable configuration extends through a finite prefix to a cut; all included operations after the cut are ordered after the removed prefix.}
\label{fig:goodcut}
\end{figure*}

\begin{theorem}[Log restoration theorem]
Let $A$ be a wait-free eventually linearizable implementation of an $n$-process log from linearizable base objects $O$.  Then there exist a reachable configuration $C_0$ of $A$ and a finite log prefix $P_0$ such that the following shifted implementation $A^{\star}$ is wait-free and linearizable: initialize the base objects and process local states as in $C_0$; encode each external append tag by a fresh internal tag under a fixed computable injection $\rho$ whose range is disjoint from $P_0$ and whose inverse on that range is computable; run $A$ from that configuration; when $A$ returns $P_0\cdot R$, return $\rho^{-1}(R)$.
\end{theorem}

\begin{proof}
Choose a stable node $C$ by the stable-node lemma, and then choose $C_0$, $P_0$, and the prefix $\alpha_0$ from the good-cut lemma.  Wait-freedom is inherited from $A$.  Consider any execution $\beta$ of $A^{\star}$.  There is a corresponding execution $\alpha_0\beta'$ of $A$, where every append tag in $\beta'$ is renamed by $\rho$ and responses are not quotiented by $P_0$.  Since $C$ was stable and $\alpha_0$ extends $C$, the history $\alpha_0\beta'$ is $t$-linearizable.  By the good-cut lemma, all operations represented by $P_0$ precede every operation of $\beta'$ that appears in this linearization.  Therefore every complete operation in $\beta'$ returns a prefix of the form $P_0\cdot R$.

Remove from the $t$-linearization all operations represented by $P_0$, and apply $\rho^{-1}$ to every remaining operation tag and to every tag in each remaining response.  The result is a legal sequential log history for the operations in $\beta$.  Real-time order in $\beta$ is respected because all events of $\beta'$ occur after $t$.  Hence the transformed sequential history is a linearization of $\beta$.

Finally, $\alpha_0$ is finite, so it accesses only finitely many members of any standard unbounded base-object supply.  Consequently only finitely many base objects can be in nonstandard reachable states in $C_0$; all untouched objects retain their standard initial states.
\end{proof}

The theorem explains why the removed prefix must be $P_0$, the post-cut state.  If $o_0$ returned only $L_0$, then $P_0=L_0\cdot x_0$.  Quotienting by $L_0$ alone leaves $x_0$ as the first apparent element of the fresh log.

\section{Hierarchy consequences}\label{sec:hierarchy}

The restoration theorem immediately recovers the classical consensus hierarchy for target-side eventual logs.

\begin{theorem}[Target-side exact hierarchy]
For every state-robust type $T$, $\lelog(T)=c(T)$.
\end{theorem}

\begin{proof}
For the lower bound, let $n\le c(T)$.  By the definition of $c(T)$, linearizable $T$ objects and registers implement $n$-process consensus.  Use an independent copy of that implementation for each consensus instance in Herlihy's universal construction.  The resulting standard class-level supply of linearizable $T$ objects and registers implements a linearizable $n$-process log.  A linearizable log is eventually linearizable, so $\lelog(T)\ge n$.

For the upper bound, suppose a standard supply of linearizable $T$ objects and registers implements an eventually linearizable $n$-process log.  By the log restoration theorem, the shifted supply implements a linearizable log and hence solves $n$-process consensus.  The finite prefix reaching the shifted configuration accesses only finitely many $T$ objects, so only finitely many copies start in nonstandard reachable states.  Fixed process-local states can be hardwired, and fixed register states can be represented by a value translation.  State robustness therefore gives an $n$-consensus protocol from a standard supply of $T$ objects and registers, so $n\le c(T)$.
\end{proof}

\begin{figure*}[t]
\centering
\begin{tikzpicture}[font=\footnotesize,>=Latex]
  \node[paper box,minimum width=2.1cm,minimum height=0.82cm] (power) at (1.2,0)
    {$c(T)\ge n$};
  \node[paper box,minimum width=2.5cm,minimum height=0.82cm] (linlog) at (4.6,0)
    {linearizable\\$\Log_n$};
  \node[paper key,minimum width=2.5cm,minimum height=0.82cm] (evlog) at (8.2,0)
    {eventual\\$\Log_n$};
  \node[paper box,minimum width=3.4cm,minimum height=0.82cm] (alltypes) at (12.8,0)
    {every computable deterministic\\eventual target};

  \draw[paper arrow] (power) -- node[above]{Herlihy} (linlog);
  \draw[paper arrow] (linlog) -- node[above]{weaken} (evlog);
  \draw[paper arrow] (evlog) -- node[above]{replay} (alltypes);
  \draw[paper transform] (evlog.north) -- ++(0,0.82) -|
    node[pos=0.45,above]{restoration + first log entry} (power.north);
\end{tikzpicture}
\caption{The target-side equality.  Classical universality reaches an eventual log from consensus power; restoration and the first log entry close the loop back to $n$-consensus.  Replay supplies every computable deterministic target.}
\Description{A loop connects consensus power, a linearizable log, and an eventual log; a replay branch from the eventual log reaches every computable eventual type.}
\label{fig:hierarchy}
\end{figure*}

\begin{corollary}
Registers have $\lelog=1$; fetch-and-increment and fetch-and-add have $\lelog=2$; and any type of infinite consensus number has infinite $\lelog$, provided it is state-robust.
\end{corollary}

Combining the target-side theorem with universality shows that $\lelog$ is a genuine universal hierarchy number, in the same sense that Herlihy's consensus number classifies universal implementability.

\begin{corollary}[Universal characterization of $\lelog$]
For every state-robust type $T$ and every $n$, the following are equivalent: (i) $\lelog(T)\ge n$; (ii) linearizable objects of type $T$ and registers wait-free implement an eventually linearizable $n$-process object of every computable deterministic $n$-process type; (iii) $n\le c(T)$.
\end{corollary}

\begin{proof}
(i)$\Rightarrow$(ii): by definition $\lelog(T)\ge n$ yields an eventually linearizable $n$-process log, which by the universality theorem implements any computable deterministic $n$-process type.  (ii)$\Rightarrow$(i): a log is itself a computable deterministic $n$-process type.  (i)$\Leftrightarrow$(iii): the target-side hierarchy theorem.
\end{proof}

\begin{remark}
Like the consensus number, $\lelog$ classifies synchronization power, not every behavioral feature of an object.  It answers the universal question for computable deterministic eventually linearizable objects; it need not decide every pairwise reduction between types.
\end{remark}

For the black-box interpretation, where base objects are themselves specified only by eventual linearizability, the restoration theorem gives a one-sided result.

\begin{theorem}[General upper bound for eventual bases]\label{thm:elog-upper}
For every state-robust type $T$, $\elog(T)\le c(T)$.
\end{theorem}

\begin{proof}
Assume an implementation $B$ uses eventually linearizable $T$ objects and registers to implement an eventually linearizable $n$-process log.  The implementation must be correct for all allowed histories of the base objects.  In particular, it is correct in executions where every base object of type $T$ happens to be linearizable, because linearizability is a special case of eventual linearizability.  Thus linearizable $T$ objects and registers implement an eventually linearizable $n$-process log.  The previous theorem gives $n\le c(T)$.
\end{proof}

The equality for $\lelog$ does not supply a lower bound for $\elog$.  Herlihy's construction may consume an unbounded sequence of fresh linearizable instances, whereas $\elog$ permits only a fixed finite pool of eventual instances; moreover, finitely many early anomalies can permanently corrupt a classical implementation's internal state.  Thus exact eventual-base lower bounds require either a self-stabilizing construction or a type-specific robustness proof.  Keeping the two resource conventions explicit lets us state the target-side hierarchy as a theorem while treating eventual-base lower bounds with the care they require.

The next two sections give a general finite-contamination compiler and use it to obtain matching lower bounds for both standard long-lived counters.

\paragraph{One-shot contrast.}
Call a one-shot type \emph{solo-explainable} if it has a wait-free register implementation in which every response has a weak-consistency explanation.  Such an implementation is eventually linearizable because a finite-process one-shot history has only finitely many responses.  If a solo-explainable type $T$ had $\elog(T)\ge2$, composition would give an eventual two-process log from registers; the restoration theorem would then give a linearizable two-process log from registers, contradicting the impossibility of two-process consensus.  Hence $\elog(T)=1$.  For consensus, each process publishes its proposal and returns the lowest-index published value; for test-and-set, a process reads a bit and, if it is zero, writes one and wins.  Both register algorithms are wait-free and solo-explainable, so both types have eventual log number $1$.

\section{Stabilizing predecessor certificates}\label{sec:certificates}

We isolate the ordering argument used by our counter constructions.  An operation $x$ first announces a unique tag, computes a finite set $C_x$ of predecessor tags, writes the single-assignment register $\Cert[x]:=C_x$, then reads $\Cert[y]$ once for every $y\in C_x$.  It returns the tags in the order $\StableOrder(C_x)$ defined below.

\begin{definition}[Stabilizing certificate execution]\label{def:certificates}
A two-process certificate execution is \emph{safe} if every completed $x$ has a finite $C_x$ such that $x\notin C_x$, every member was invoked before $x$ responds, and $C_x$ contains every completed operation of $x$'s process that precedes $x$.  It is \emph{stabilizing} if there are a set $M$ of invoked operations containing every completed operation, a total order $\prec$ of $M$ in which every element has finitely many predecessors, and a finite set $B\subseteq M$ such that every $x\in M\setminus B$ completes its certificate scan and: (i) $C_x=\{y\in M:y\prec x\}$; and (ii) that scan observes at most one unset certificate.  Pending operations may be omitted from $M$; any included operation that does not finish its scan belongs to $B$.
\end{definition}

For a finite set $A$, form a directed graph on $A$.  Add an edge from each earlier to each later operation of the same process.  For distinct $a,b\in A$, add $a\to b$ if the observed $\Cert[b]$ is set and contains $a$, or if the observed $\Cert[a]$ is set and does not contain $b$.  Collapse strongly connected components, topologically sort the component DAG with a fixed tie-breaker, and inside each component use a fixed global order that increases with every process's local sequence number.  This total order is $\StableOrder(A)$; Figure~\ref{fig:certificate-compiler} shows the invariant established below.

\begin{figure*}[t]
\centering
\begin{tikzpicture}[font=\footnotesize,>=Latex]
  \node[paper frame,minimum width=5.7cm,minimum height=2.45cm] (stage1) at (2.85,0) {};
  \node[font=\small] at (2.85,0.92) {\textbf{1. Certificate graph}};
  \node[paper key,minimum width=2.15cm,minimum height=0.88cm] (P) at (1.45,-0.18)
    {$P$\\finite, possibly cyclic};

  \node[circle,draw,inner sep=1.2pt] (q1) at (3.5,-0.18) {$q_1$};
  \node[circle,draw,inner sep=1.2pt] (q2) at (4.35,-0.18) {$q_2$};
  \node[circle,draw,inner sep=1.2pt] (q3) at (5.2,-0.18) {$q_3$};
  \node at (4.35,0.35) {$Q_x$};
  \draw[paper transform] (P.east) -- (q1.west);
  \draw[paper arrow] (q1) -- (q2);
  \draw[paper arrow] (q2) -- (q3);
  \node at (3.2,-0.9) {all cross edges point $P\to Q_x$};

  \node[paper frame,minimum width=4.25cm,minimum height=2.45cm] (stage2) at (9.2,0) {};
  \node[font=\small] at (9.2,0.92) {\textbf{2. Condensation DAG}};
  \node[paper key,minimum width=1.2cm,minimum height=0.68cm] (Pc) at (7.8,-0.18) {$[P]$};
  \node[circle,draw,inner sep=1.2pt] (r1) at (9.05,-0.18) {$q_1$};
  \node[circle,draw,inner sep=1.2pt] (r2) at (10.0,-0.18) {$q_2$};
  \node[circle,draw,inner sep=1.2pt] (r3) at (10.95,-0.18) {$q_3$};
  \draw[paper arrow] (Pc) -- (r1);
  \draw[paper arrow] (r1) -- (r2);
  \draw[paper arrow] (r2) -- (r3);

  \node[paper frame,minimum width=3.25cm,minimum height=2.45cm] (stage3) at (14.3,0) {};
  \node[font=\small] at (14.3,0.92) {\textbf{3. Returned prefix}};
  \node at (14.3,0.38) {$\StableOrder(C_x)$};
  \node[paper key,minimum width=0.68cm,minimum height=0.58cm] (outp) at (13.3,-0.38) {$P_0$};
  \node[paper box,minimum width=0.58cm,minimum height=0.58cm] (out1) at (14.0,-0.38) {$q_1$};
  \node[paper box,minimum width=0.58cm,minimum height=0.58cm] (out2) at (14.7,-0.38) {$q_2$};
  \node[paper box,minimum width=0.58cm,minimum height=0.58cm] (out3) at (15.4,-0.38) {$q_3$};

  \draw[paper transform] (stage1.east) -- node[above]{collapse SCCs} (stage2.west);
  \draw[paper transform] (stage2.east) -- node[above]{topological sort} (stage3.west);
\end{tikzpicture}
\caption{The certificate compiler.  Positional closure confines anomalies to $P$; SCC condensation replaces that subgraph by $[P]$, and topological sorting expands it with one fixed order $P_0$ while preserving the clean suffix.}
\Description{Three stages show a cyclic contaminated prefix and clean chain, their SCC condensation DAG, and the final returned prefix with a fixed reordered prefix followed by the unchanged clean suffix.}
\label{fig:certificate-compiler}
\end{figure*}

\begin{theorem}[Certificate compiler]\label{thm:compiler}
Any wait-free safe stabilizing certificate implementation yields a wait-free eventually linearizable two-process log.
\end{theorem}

\begin{proof}
Wait-freedom follows because each certificate and scan is finite.  Safety gives weak consistency: $\StableOrder(C_x)\cdot x$ contains only previously invoked operations, excludes $x$ from its returned prefix, respects process order, includes all earlier completed operations of $x$'s process, and is a legal sequential log explanation.  A finite history is handled by choosing a cut after its last event, so assume the history is infinite.

Fix $M,\prec,B$.  Let $Z$ contain $B$ and every member of $M$ named by a published certificate of an operation in $B$, and let $P$ be the shortest initial segment of $\prec$ containing $Z$.  It is finite.  If $p\in P\setminus B$, exactness gives $C_p\cap M\subseteq P$; if $p\in B$, the definition of $P$ gives the same inclusion.  Thus no available prefix certificate names a vertex of $M\setminus P$.

Choose a time $T^\ast$ after every operation in $P$ has been invoked and every $\Cert[p]$, $p\in P$, has either been written or is permanently unset.  Choose a high-level cut $t$ after every operation invoked before $T^\ast$ that ever completes has responded.  A completed operation $x$ responding after $t$ lies outside $P$ and has
\[
 C_x=P\cup Q_x,\qquad Q_x=\{q\notin P:q\prec x\}.
\]
For $p\in P$ and $q\in Q_x$, if $\Cert[q]$ is available then it contains $p$; otherwise the one-unavailable condition makes $\Cert[p]$ available and it excludes $q$.  Hence the certificate graph has $p\to q$ and no reverse edge.  Program-order edges cannot point back: if $q$ completed before $p$ was invoked, exactness would put $p$ before $q$ in $\prec$ while safety of $C_q$ would require the not-yet-invoked $p$ to belong to $C_q$, a contradiction.  The same argument, together with the certificate rule, shows that for $u\prec v$ in $Q_x$ the graph has $u\to v$ and no reverse edge.  The graph induced by $P$ is fixed after $t$, so its SCC rule yields one fixed permutation $P_0$.  Therefore every operation responding after $t$ returns
\[
 P_0\cdot Q_x^\prec .
\]
The order $P_0$ followed by $M\setminus P$ in $\prec$ order is a $t$-linearization; pending members of $M$ are completed with the responses prescribed by their positions.  Exactness and safety imply suffix real time: if $x$ responds before $y$ is invoked, then $y\prec x$ would put the not-yet-invoked $y$ in $C_x$.  Thus the compiled log is eventually linearizable.
\end{proof}

\paragraph{Predecessor-decodable updates.}
Call a type predecessor-decodable if it has fresh coded updates $U(c)$ and a computable decoder such that, in every legal sequential execution using distinct codes, the response to $U(c)$ decodes exactly the finite set of earlier codes.  One eventual object of this type generates safe certificates by announcing $c$, invoking $U(c)$, decoding the response, and freezing the resulting set before publishing it.  Weak consistency gives safety for every response; an eventual linearization gives one common order and only finitely many inexact certificates.  Put the at most two included high-level operations that never finish their certificate scans into the finite exceptional set.  For a clean $x$, its frozen set contains no operation invoked after the base response of $x$; at that moment the peer has at most one unfinished operation.  Later peer operations therefore cannot create a second unavailable member during the non-atomic scan.  The compiler gives the following.

\begin{corollary}[Decodable-update lower bound]\label{cor:decodable}
Every wait-free predecessor-decodable type $T$ has $\elog(T)\ge2$.  If $T$ is state-robust and $c(T)=2$, then $\elog(T)=2$.
\end{corollary}

\begin{proof}
Fix an eventual-$T$ history and a $\tau$-linearization $S_T$ of its coded updates.  Let $M$ be the corresponding invoked high-level operations in $S_T$, ordered as there; it contains every completed operation.  Put into $B$ every member whose update invocation is among the first $\tau$ base events and every member whose high-level operation never finishes its certificate scan.  The first set is finite, and well-formedness bounds the second by two.

For every response, base weak consistency gives a legal update explanation.  Decoding it yields a finite set of previously announced operations, excludes the current operation, and includes all completed predecessors of its process, so certificates are safe.  For $x\in M\setminus B$, clause~(iv) makes the decoded set exactly the predecessors of $x$ in $S_T$.  Freeze this set when the update returns.  At that moment the peer has at most one unfinished operation; any peer operation invoked later cannot belong to the frozen exact set.  Hence the subsequent non-atomic scan observes at most one unavailable certificate.  Definition~\ref{def:certificates} and Theorem~\ref{thm:compiler} give $\elog(T)\ge2$, and Theorem~\ref{thm:elog-upper} gives equality when $c(T)=2$ and $T$ is state-robust.
\end{proof}

Zero-initialized unbounded old-value fetch-and-add is a direct instance: an injective computable encoding assigns operation $x$ the fresh power $2^{e(x)}$, and the returned integer decodes to its set bits.  The same lower-bound argument applies to unbounded fetch-and-OR with fresh bits and, more generally, to effective square-free accumulators whose finite products of fresh generators are uniquely decodable.

\begin{figure*}[t]
\centering
\begin{tikzpicture}[font=\footnotesize,>=Latex]
  \node[font=\small] at (1.85,1.95) {\textbf{Source-specific identity recovery}};
  \draw[gray!55] (-1.5,1.68) -- (5.2,1.68);
  \node[font=\small] at (7.25,1.05) {\textbf{Shared invariant}};
  \node[font=\small] at (10.65,1.05) {\textbf{Order}};
  \node[font=\small] at (13.75,1.05) {\textbf{Log}};

  \node[paper box,minimum width=3.0cm,minimum height=1.12cm] (faa) at (0,0.9)
    {identity-bearing update\\$\FAA(2^{e(x)})$};
  \node[paper box,minimum width=3.0cm,minimum height=1.12cm] (fai) at (0,-0.9)
    {rank-only update\\$\FAI()\mapsto m$};

  \node[paper box,minimum width=3.0cm,minimum height=1.12cm] (decode) at (3.7,0.9)
    {decode set bits\\$\bits(m)$};
  \node[paper box,minimum width=3.0cm,minimum height=1.12cm] (lift) at (3.7,-0.9)
    {announcements + one hole\\cardinality lift};

  \node[paper key,minimum width=3.15cm,minimum height=1.18cm] (cert) at (7.25,0)
    {safe stabilizing\\certificate $C_x$};
  \node[paper box,minimum width=2.75cm,minimum height=1.18cm] (compiler) at (10.65,0)
    {SCC certificate\\compiler};
  \node[paper box,minimum width=2.45cm,minimum height=1.18cm] (log) at (13.75,0)
    {eventual\\$\Log_2$};
  \node[paper box,minimum width=3.2cm,minimum height=0.9cm] (target) at (13.75,-1.65)
    {every computable\\deterministic target};

  \draw[paper arrow] (faa) -- (decode);
  \draw[paper arrow] (fai) -- (lift);
  \draw[paper arrow] (decode.east) -- ++(0.45,0) |- (cert.north west);
  \draw[paper arrow] (lift.east) -- ++(0.45,0) |- (cert.south west);
  \draw[paper arrow] (cert) -- (compiler);
  \draw[paper arrow] (compiler) -- (log);
  \draw[paper arrow] (log) -- node[right]{replay} (target);
\end{tikzpicture}
\caption{Two certificate generators share one compiler.  Self-describing updates decode identities directly; FAI instead lifts a rank using the one-hole rule.  Both routes produce the same stabilizing certificate, eventual log, and universal replay target.}
\Description{Parallel FAA and FAI paths generate stabilizing predecessor certificates, which flow through one SCC compiler to an eventual two-process log and universal replay.}
\label{fig:certificate-pipeline}
\end{figure*}

Figure~\ref{fig:certificate-pipeline} separates the source-specific task of recovering predecessor identities from the source-independent task of ordering them.

\section{Rank lifting for eventual counters}\label{sec:counters}

Predecessor-decodability is sufficient but not necessary.  Fetch-and-increment returns only a rank, not predecessor identities.  With two processes, announcements and cardinality recover the identities because at most one announced operation can be stalled between obtaining and publishing its ticket.

We use one wait-free eventually linearizable unbounded $\FAI$ object $F$, initially $0$.  For $i\in\{0,1\}$, $\Head[i]$ is a single-writer register initially $-1$, and $\Tag[i,k]$, $\Ticket[i,k]$, and $\Cert[i,k]$ are single-assignment registers initially unset.

\begin{quote}
\textbf{Rank lifting from one eventual FAI.}
For process $p_i$ to append its $k$-th tag $x=(i,k,\mathit{payload})$:
\begin{enumerate}[leftmargin=2em]
    \item write $\Tag[i,k]:=x$ and then $\Head[i]:=k$;
    \item let $m:=F.\FAI()$ and write $\Ticket[i,k]:=m$;
    \item read $h:=\Head[1-i]$ once and then read $\Ticket[1-i,j]$ for $0\le j\le h$;
    \item let $C$ contain all $(i,j)$ with $j<k$ and every scanned $(1-i,j)$ whose observed ticket is set and smaller than $m$;
    \item if $|C|=m$, set $A:=C$; else, if $|C|=m-1$ and exactly one scanned ticket is unset, add its operation to obtain $A$; otherwise set $A:=\{(i,j):j<k\}$;
    \item write $\Cert[i,k]:=A$, scan $\Cert[a]$ and read $\Tag[a]$ for $a\in A$, and return those tags in $\StableOrder(A)$.
\end{enumerate}
\end{quote}

Figure~\ref{fig:fai-hole} isolates the cardinality test that distinguishes the two otherwise ambiguous pause points.

\begin{figure*}[t]
\centering
\begin{tikzpicture}[font=\footnotesize,>=Latex]
  \node[font=\small] at (3.7,1.48) {\textbf{(a) Peer pauses before FAI}};
  \node[anchor=east] at (1.1,0.48) {peer $q$};
  \draw[paper secondary] (1.3,0.48) -- (3.52,0.48);
  \draw[paper guide] (3.62,0.48) -- (5.75,0.48);
  \node[paper box,minimum width=1.55cm,minimum height=0.55cm] (la) at (2.15,0.48) {announce};
  \draw[line width=0.7pt] (3.52,0.22) -- (3.52,0.74);
  \draw[line width=0.7pt] (3.62,0.22) -- (3.62,0.74);
  \node at (3.57,0.9) {pause};
  \node[anchor=east] at (1.1,-0.28) {caller $x$};
  \draw[paper secondary] (1.3,-0.28) -- (6.65,-0.28);
  \node[paper box,minimum width=1.85cm,minimum height=0.55cm] (lx) at (4.55,-0.28) {$\FAI()\mapsto m$};
  \draw[paper guide] (5.75,-0.62) -- (5.75,0.82);
  \node at (5.75,0.98) {scan};
  \node[paper key,minimum width=4.75cm,minimum height=0.82cm] at (3.75,-1.28)
    {$|C|=m$\\\textbf{exclude }$q$};

  \node[font=\small] at (11.25,1.48) {\textbf{(b) Peer pauses after FAI}};
  \node[anchor=east] at (8.65,0.48) {peer $q$};
  \draw[paper secondary] (8.85,0.48) -- (12.32,0.48);
  \draw[paper guide] (12.42,0.48) -- (14.45,0.48);
  \node[paper box,minimum width=1.55cm,minimum height=0.55cm] (ra) at (9.65,0.48) {announce};
  \node[paper box,minimum width=1.15cm,minimum height=0.55cm] (rf) at (11.25,0.48) {FAI};
  \draw[line width=0.7pt] (12.32,0.22) -- (12.32,0.74);
  \draw[line width=0.7pt] (12.42,0.22) -- (12.42,0.74);
  \node at (12.37,0.9) {pause};
  \node[anchor=east] at (8.65,-0.28) {caller $x$};
  \draw[paper secondary] (8.85,-0.28) -- (14.8,-0.28);
  \node[paper box,minimum width=1.85cm,minimum height=0.55cm] (rx) at (13.15,-0.28) {$\FAI()\mapsto m$};
  \draw[paper guide] (14.45,-0.62) -- (14.45,0.82);
  \node at (14.45,0.98) {scan};
  \node[paper key,minimum width=4.95cm,minimum height=0.82cm] at (11.4,-1.28)
    {$|C|=m-1$\\\textbf{include }$q$};
\end{tikzpicture}
\caption{Rank lifting distinguishes the hole by its pause location: before the peer's FAI the rank already matches $|C|$, while after that FAI the rank exceeds $|C|$ by one.}
\Description{Two cases compare an announced peer operation that stalls before versus after its FAI.  The returned rank and predecessor-set cardinality determine whether to exclude or include the peer operation.}
\label{fig:fai-hole}
\end{figure*}

\begin{lemma}[Rank lifting]\label{lem:rank-lifting}
The construction above is a wait-free safe stabilizing certificate implementation.
\end{lemma}

\begin{proof}
\emph{Wait-freedom and safety.}
The frozen value $h$ bounds the scan, and all other work is finite.  Every member of $A$ was announced before $x$ responds.  All own predecessors are included unconditionally, including in the fallback, so every certificate is safe.

\emph{Visibility and one ticket hole.}
Map each high-level operation to its unique FAI call.  Fix a $\tau$-linearization $S_F$ of the base history and number its operations from $0$.  If $q<_{S_F}x$ and $x$'s FAI invocation is after $\tau$, then $q$'s announcement is visible when $x$ reads the corresponding head: an invocation of $q$ after the response of $x$ would, after $\tau$, force $x<_{S_F}q$.  Every previous high-level operation of $x$'s process has completed and published its ticket.  The other process has at most one incomplete high-level operation, so the frozen scan observes at most one unset ticket.

\emph{Finite ticket contamination.}
Let $D$ be the operations in $S_F$ whose FAI invocation occurs among the first $\tau$ base events, and let $U$ be the operations in $S_F$ whose high-level append never finishes its certificate scan.  Both sets are finite; well-formedness gives $|U|\le2$.  Choose $R\ge-1$ at least every $S_F$-position of a member of $D\cup U$ and every ticket eventually published by a member of $D$.  Define
\[
 B=\{x\in S_F:\operatorname{pos}_{S_F}(x)\le R\}.
\]
This is finite even when $S_F$ itself is finite.  An operation outside $B$ is invoked after $\tau$ and finishes its scan, so its returned ticket equals its zero-based position in $S_F$ and exceeds $R$.  Every published ticket of an operation in $B$ is at most $R$: it is either an early value included in the choice of $R$, or the exact position of a post-$\tau$ operation.

\emph{Cardinality reconstruction.}
Let $x\notin B$ have ticket $m$.  Every predecessor of $x$ whose ticket is published is placed in $C$: a member of $B$ has ticket at most $R<m$, while a clean suffix operation has ticket equal to its position.  Every own high-level predecessor also precedes $x$ in $S_F$: if its FAI invocation is early it belongs to $D\subseteq B$, and otherwise its response and $x$'s invocation are both after $\tau$, so suffix real time orders it first.  Conversely every peer operation placed in $C$ either belongs to the initial segment $B$ or has a clean ticket equal to a position below $m$, and therefore precedes $x$.

By visibility, the only possible missing predecessor is the unique scanned operation with an unset ticket.  Since a zero-initialized FAI operation at position $m$ has exactly $m$ predecessors, either $|C|=m$ and $C=\Pred_{S_F}(x)$, or $|C|=m-1$ and adjoining that unique hole gives $\Pred_{S_F}(x)$.  These two cases also cover every pending call: if its FAI is included before $x$ in $S_F$, the rank counts it and the second branch adds it; if it is omitted or ordered after $x$, all $m$ true predecessors are already in $C$ and the first branch excludes it.  An operation announced but stalled before invoking FAI is handled by the same first branch.  Thus the fallback is never taken outside the finite prefix $B$.

\emph{Certificate availability.}
For a clean $x$, every own predecessor has completed and published its certificate.  At most the other process's current operation can be a predecessor without a published certificate.  Hence the actual ordering scan observes at most one unavailable certificate.  Taking $M$ to be the operations of $S_F$ and $\prec$ its order proves stabilization.
\end{proof}

\begin{theorem}[Exact eventual counter values]\label{thm:counters}
For standard zero-initialized unbounded old-value fetch-and-increment and fetch-and-add,
\[
   \elog(\FAI)=\elog(\FAA)=2 .
\]
\end{theorem}

\begin{proof}
The compiler and Lemma~\ref{lem:rank-lifting} give $\elog(\FAI)\ge2$.  An FAA object restricted to increments of $1$ supplies an FAI history, so $\elog(\FAA)\ge2$ as well; alternatively FAA instantiates Corollary~\ref{cor:decodable}.  Both types are state-robust and have consensus number $2$, so Theorem~\ref{thm:elog-upper} gives the matching upper bounds.
\end{proof}

\paragraph{Boundaries.}
The result uses zero-initialized unbounded counters.  With an unknown offset $q$, the $r$-th FAI response is $q+r$ and the cardinality test fails; fixed-width wraparound likewise destroys eventual rank uniqueness.  The rank lift is intrinsically two-process: with three processes, two peers may expose two ticket holes, and a rank can reveal how many but not which operations precede the caller.  This matches the upper bound $\elog(\FAI)\le c(\FAI)=2$.

\section{Conclusion}

Table~\ref{tab:summary} collects the resulting hierarchy.  One-shot consensus and test-and-set collapse under the finite-pool eventual-base convention, whereas both standard long-lived counters retain their classical level $2$.  The FAI result is the nontrivial case: rank does not encode predecessor identities, but two-process one-hole cardinality reconstructs them.

\begin{table}[t]
\centering
\caption{Consensus number, class-level linearizable-base $\lelog$, and finite-pool eventual-base $\elog$ for the types treated in this paper.}
\label{tab:summary}
\begin{tabular}{@{}lccc@{}}
\toprule
Type $T$ & $c(T)$ & $\lelog(T)$ & $\elog(T)$ \\
\midrule
Registers                & $1$        & $1$        & $1$ \\
Consensus                & $\infty$   & $\infty$   & $1$ \\
Test-and-set             & $2$        & $2$        & $1$ \\
Fetch-and-increment      & $2$        & $2$        & $2$ \\
Fetch-and-add            & $2$        & $2$        & $2$ \\
$\Log_n$                 & $n$        & $n$        & $n$ \\
\bottomrule
\end{tabular}
\end{table}

Eventually linearizable logs are the universal objects for eventual shared-memory implementations.  Their finite prefix can be removed from a reachable configuration when bases are linearizable, yielding the exact class-level hierarchy $\lelog(T)=c(T)$ and the finite-pool upper bound $\elog(T)\le c(T)$.  For lower bounds with eventual bases, stabilizing certificates confine all inconsistent ordering information to one finite log prefix.  Directly decodable updates such as FAA generate predecessor identities; FAI instead lifts ranks to identities using announcements and the fact that two processes create at most one unresolved ticket.  This gives the exact black-box values $\elog(\FAI)=\elog(\FAA)=2$.

\end{document}